\documentclass{iopart}
\usepackage{iopams}
\usepackage{graphicx}
\begin{document}

\title[Possibility of field-induced IC order in quasi-1D frustrated spin system]{
Possibility of field-induced incommensurate order in quasi-one-dimensional frustrated spin system%
}


\author{N Maeshima$^{1,2}$, K Okunishi$^3$, K Okamoto$^4$, T Sakai$^5$ and K Yonemitsu$^{2,6}$}

\address{$^1$Department of Chemistry, Tohoku University, Aramaki, Aoba-ku, Sendai 980-8578, Japan}

\address{$^2$Institute for Molecular Science, Okazaki 444-8585, Japan}

\address{$^3$Department of Physics, Niigata University, Igarashi 2, Niigata 950-2181, Japan}

\address{$^4$Department of Physics, Tokyo Institute of Technology, Meguro-ku, Tokyo 152-8551, Japan}

\address{$^5$Japan Atomic Energy Agency, SPring-8, Sayo, Hyogo 679-5148, and CREST JST, Japan}


\address{$^6$Graduate University for Advanced Studies, Okazaki 444-8585, Japan}

\ead{maeshima@ims.ac.jp}

\begin{abstract}
We study an incommensurate long-range order induced by an external magnetic field in a quasi-one-dimensional bond-alternating spin system,  F$_5$PNN,  focusing on the role of the frustrating interaction which can be enhanced by a high-pressure effect. On the basis of the density matrix renormalization group analysis of a microscopic model for F$_5$PNN, we present several $H$-$T$ phase diagrams for typical parameters of the frustrating next-nearest-neighbour coupling and the interchain interaction, and  then discuss how the field-induced incommensurate order develops by the frustration effect in such phase diagrams. A magnetization plateau at half the saturation moment is also mentioned. 
\end{abstract}

\pacs{75.10.Jm, 75.40.Cx, 75.50.Ee}

\submitto{\JPCM}

\maketitle

\section{Introduction}

Quasi-one-dimensional quantum spin systems exhibit several interesting field-induced phenomena.
In a gapped spin chain  an external magnetic field exhibits a quantum 
phase transition to the gapless Tomonaga-Luttinger liquid phase 
at some critical field $H_{c1}$~\cite{sakai1,sakai11}.
The Tomonaga-Luttinger liquid phase basically continues up to the saturation field $H_{c2}$ and is characterized by the power-law decay of the spin correlation function:  
\begin{eqnarray}
&&\langle S_0^x S_r^x \rangle \sim (-1)^r r^{-\eta _x}, \label{sx}  \\
&&\langle S_0^z S_r^z \rangle -m^2 \sim \cos(2k_F r) r^{-\eta_z},
\label{sz}
\end{eqnarray}
where $m$ is the magnetization along the field $H \parallel z$, and the Fermi wave number $k_F$ depends on the magnetization. 
Since these exponents satisfy $\eta_x < \eta_z$ in usual antiferromagnets, the commensurate antiferromagnetic long-range order perpendicular to $H$ should appear in the presence of interchain interactions.
In fact, such a field-induced antiferromagnetic order has been theoretically proposed~\cite{sakai2} and experimentally observed in some quasi-one-dimensional gapped systems~\cite{honda,tanaka}.
Here it should be noted that the ordered phase can be also interpreted as a result of the magnon Bose-Einstein condensation~\cite{nikuni}.

In some one-dimensional (1D) frustrated systems, however, the $\eta$-inversion, namely $\eta_x > \eta_z$, has been predicted for some intermediate strength of the magnetic field between $H_{c1}$ and $H_{c2}$~\cite{suga1,suga2,suga3,chitra,mila}.  Indeed, such  $\eta$-inversion has been suggested experimentally by the NMR relaxation rate $1/T_1$ of an organic $S=1/2$ bond-alternating chain, pentafulorophenyl nitronyl nitroxide (F$_5$PNN), in a certain temperature region where a power-law behaviour of $1/T_1$ can be observed~\cite{izumi,goto}.
A more interesting point associated with the $\eta$-inversion is that the field-induced long-range {\it incommensurate order} can appear, reflecting the enhancement of the incommensurate correlation~\cite{maeshima1}.
Although the recent specific heat measurement~\cite{yoshida1,yoshida2} indicates that the field-induced commensurate antiferromagnetic order may be realized for F$_5$PNN, the NMR experiments are still suggesting that an enhancement of the incommensurate correlation in F$_5$PNN.  Thus, it seems that F$_5$PNN is located in the vicinity of the boundary between the commensurate order phase and the incommensurate order phase.
In this paper, we discuss the possibility of realizing the field-induced incommensurate order in F$_5$PNN, putting a special emphasize on a high-pressure effect which can enhance the frustrating interaction.
We examine how the incommensurate order appears in the phase diagram on the basis of density matrix renormalization group calculations of a microscopic model for F$_5$PNN.
We also mention a possibility for a magnetization plateau and coexistence of the commensurate and incommensurate orders. 

\section{Model}

The experimentally observed $\eta$-inversion and the field-dependent crossover of 
the bond-alternating ratio~\cite{hosokoshi} 
 have indicated the presence of frustration in F$_5$PNN. 
Thus, we consider the bond-alternating spin chain with the 
next-nearest-neighbour interaction in a magnetic field 
\begin{eqnarray}
{\hat H}=&&J_1\sum _j {\vec S}_{2j} \cdot {\vec S}_{2j+1} + 
J_2 \sum_j {\vec S}_{2j+1} \cdot {\vec S}_{2j+2} \\
&+& J'\sum_j {\vec S}_{2j} \cdot {\vec S}_{2j+2}
 - H\sum _j S_j^z, 
\label{ham}
\end{eqnarray}
as a realistic model for F$_5$PNN, where $J_1$ and $J_2$ ($\ne J_1$) denote the alternating nearest neighbour interactions, and $J'$ means the next-nearest neighbour coupling.
In what follows, $J_1$ is set to unity for simplicity.
For $H_{c1}<H<H_{c2}$, the model is basically in the  Tomonaga-Luttinger liquid phase with the Fermi wave number $k_F=(1/2-m)\pi$.

\section{Mechanism of the field-induced incommensurate order}

Let us briefly explain the mechanism of the field-induced incommensurate order.  If $J'$ is sufficiently large, the $\eta$-inversion occurs around $m\sim m_s/2$, where $m_s$ is the saturated magnetization~\cite{suga2,suga3,maeshima1}.
At the magnetic field in the $\eta$-inversion region, the dominant spin correlation should not be the usual commensurate antiferromagnetic one perpendicular to $H$, but the incommensurate one along the $H$-direction.
In the presence of interchain interactions, the long-range order corresponding to the dominant spin correlation is expected, implying that a long-range incommensurate order in the $H$  direction can be realized for the region $\eta_x > \eta_z$. 

In the present study, the dominant spin correlation is determined by two staggered susceptibilities $\chi_\perp$ and $\chi_\parallel$ corresponding to the spin correlations (\ref{sx}) and (\ref{sz}), respectively.  Low temperature behaviours of the susceptibilities are characterized by the exponents $\eta_x$ and $\eta_z$~\cite{chitra}:
\begin{equation}
\chi_\perp(T) \sim {\rm C}_\perp T^{-(2-\eta_x)} \quad {\rm and} \quad \chi_\parallel(T) \sim  {\rm C}_\parallel T^{-(2-\eta_z)},
\label{eq:diverge}
\end{equation}
where ${\rm C}_\perp$ and ${\rm C}_\parallel$ are nonuniversal coefficients.
Equation~(\ref{eq:diverge}) shows that the more dominant spin correlation corresponds to the more strongly divergent susceptibility.  
When the $\eta$-inversion occurs, for example, $\chi_\parallel$ shows stronger divergence than $\chi_\perp$.

For quantitative treatment of the field-induced order and of the accompanied phase transition, we use the above susceptibilities.
Based on the mean-field approximation for the interchain interaction $J_{\rm int}$~\cite{scalapino},
the critical temperature $T_{\rm c}$ is given by
\begin{equation}
\chi_{\gamma} (T_{\rm c}) = (zJ_{\rm int})^{-1},\label{eq:TC}
\end{equation}
where $\gamma$ is $\perp$ or $\parallel$,  and $z$ the number of adjacent chains.

\begin{figure}[btp]
\begin{center}\leavevmode
\includegraphics[width=0.95\linewidth,angle=0]{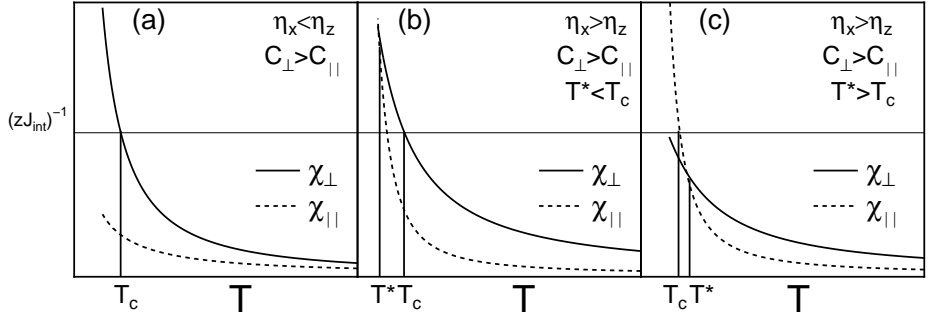}
\caption{
Schematic temperature dependence of $\chi_\perp$ and $\chi_\parallel$ in three cases. At $T=T_{\rm c}$, the larger susceptibility satisfies the relation~(\ref{eq:TC}). At $T=T^*$ in (b) and (c), $\chi_\perp$ is equal to $\chi_\parallel$.
}
\label{fig:kai}\end{center}\end{figure}

On the basis of equations~(\ref{eq:diverge}) and (\ref{eq:TC}),  we can classify  possible types of the field-induced orders; we consider the following cases, which are depicted in figure~1.
The first one is that $\eta_x<\eta_z$ [figure~\ref{fig:kai} (a)]~\cite{comment}.
In this case, since $\chi_\perp$ strongly diverges, $\chi_\perp>\chi_\parallel$ is always realized.  As $T$ decreases, $\chi_\perp$ first satisfies the relation~(\ref{eq:TC}) and the commensurate order appears below $T_{\rm c}$.
By contrast, if the $\eta$-inversion occurs (i.e. $\eta_x>\eta_z$),  $\chi_\perp<\chi_\parallel$ is naively expected. However, we should pay attention to the magnitude of the non-universal coefficients ${\rm C}_\perp$ and ${\rm C}_\parallel$.  We have found that, even for $\eta_x>\eta_z$, ${\rm C}_\perp$ is larger than ${\rm C}_\parallel$, and thus $\chi_\perp$ can be larger than $\chi_\parallel$  at high temperatures, although $\chi_\parallel$ finally exceeds $\chi_\perp$ at $T=T^*$.
If $T_{\rm c}>T^*$, therefore,  $\chi_\perp$ satisfies equation~(\ref{eq:TC}) at a higher temperature and the commensurate order appears [see figure~\ref{fig:kai} (b)].
In such a case,  another phase transition is expected to occur from the commensurate phase to the incommensurate phase at $T=T^*$.
The previous DMRG analysis of the free energy has illustrated that the transition actually occurs when $zJ_{\rm int}$ is not so large~\cite{maeshima2}.
For $T_{\rm c}<T^*$ [figure~\ref{fig:kai} (c)], $\chi_\parallel$ satisfies equation~(\ref{eq:TC}) first and only the incommensurate order appears .

\section{Phase diagrams}

By a numerical calculation of $\chi_\perp$ and $\chi_\parallel$ for a single chain and the mean-field treatment described above, we have obtained $H$-$T$ phase diagrams.
For the calculation at finite temperature in the thermodynamic limit, we use the finite temperature DMRG~\cite{wang,shibata}.
The DMRG calculation has been carried out with the maximum number of the retained bases $m=64$ and the truncation error is 
at most the order of $10^{-6}$.
Technical details to obtain $\chi_\perp$ and $\chi_\parallel$ are described in reference~\cite{maeshima1}.

In figure~\ref{fig1} we show the phase diagrams for typical three parameters of the frustrating coupling: (a) $J'=0.05$ (small $J'$ case), (b)  $J'=0.1$ (intermediate $J'$ case), and (c) $J'=0.15$ (large $J'$ case).
The interchain interaction is fixed to $zJ_{\rm int}=1/12$, and $J_2$ is fixed to a realistic value 0.45 for F$_5$PNN~\cite{maeshima1}.
In the figure, C and IC denote the commensurate antiferromagnetic-order phase and the incommensurate one, respectively.

\subsection{case (a)}
The phase diagram in the case (a) is typical for spin-gapped quasi-1D spin systems, where it should be remarked that the shape of the phase boundary between the C phase and the disordered phase looks like a semicircle.
Since $\chi_\perp$ is always larger than $\chi_\parallel$ as shown in figure~\ref{fig:kai} (a), the IC phase never appears.

\subsection{case(b)}
In the case (b), the phase diagram is clearly different from the case (a) in the following two points.
The first one is the appearance of the IC phase in a very low temperature region.  At $H=1.23$, where the magnetization $m$ at $T=0$ is equal to $m_s/2$,  the $\eta$-inversion occurs and thus $\chi_\parallel$ shows slightly stronger divergence than that of $\chi_\perp$.
Thus the IC phase can appear around $H=1.23$ in the quite low temperature region ($T<T^*$).
However, in the high temperature region ($T>T^*$), $\chi_\perp$ is larger than $\chi_\parallel$, resulting in the transition between the commensurate order phase and the disordered phase at $T=T_c$.  These behaviours of the staggered susceptibilities correspond to figure~\ref{fig:kai} (b).
Another important point is that the phase boundary between the C phase and the disordered phase is deformed from the semicircle-shape particularly around $H=1.23$.  This is because the transverse fluctuation is suppressed by the frustration effect even in $T>T^*$, resulting in the lowering of $T_{\rm c}$.

\subsection{case (c)}
In the case (c), the large $J'$ enhances the incommensurate correlation and extends the region of the IC phase.
Around $H=1.27$, $\chi_\perp$ behaves as in figure~\ref{fig:kai} (c) and thus we can observe the direct transition from the disordered phase to the IC phase, which divides the C phase into the low-$H$ part and the high-$H$ part.

\begin{figure}[tbp]
\begin{center}\leavevmode
\includegraphics[width=0.48\linewidth,angle=0]{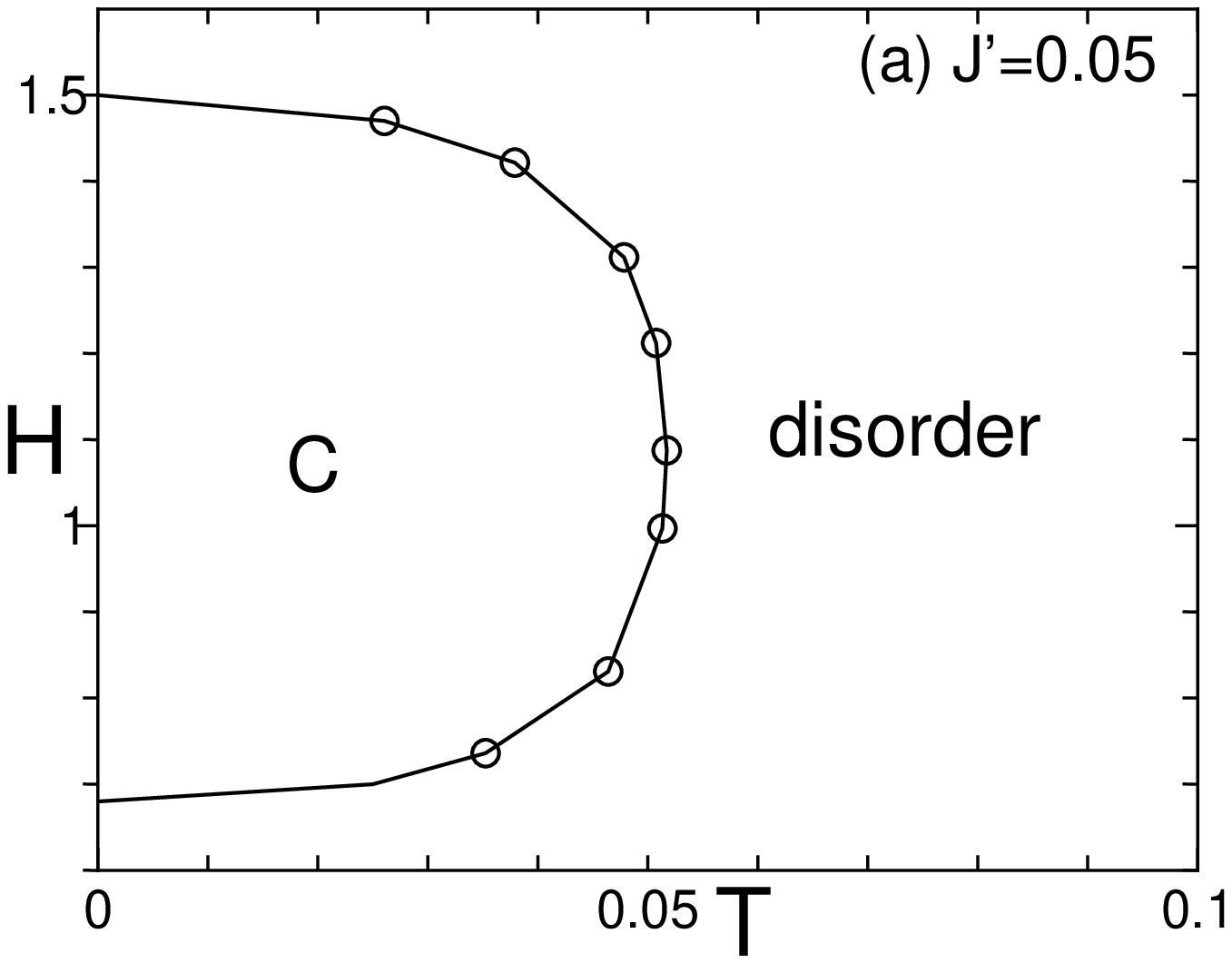}
\includegraphics[width=0.48\linewidth,angle=0]{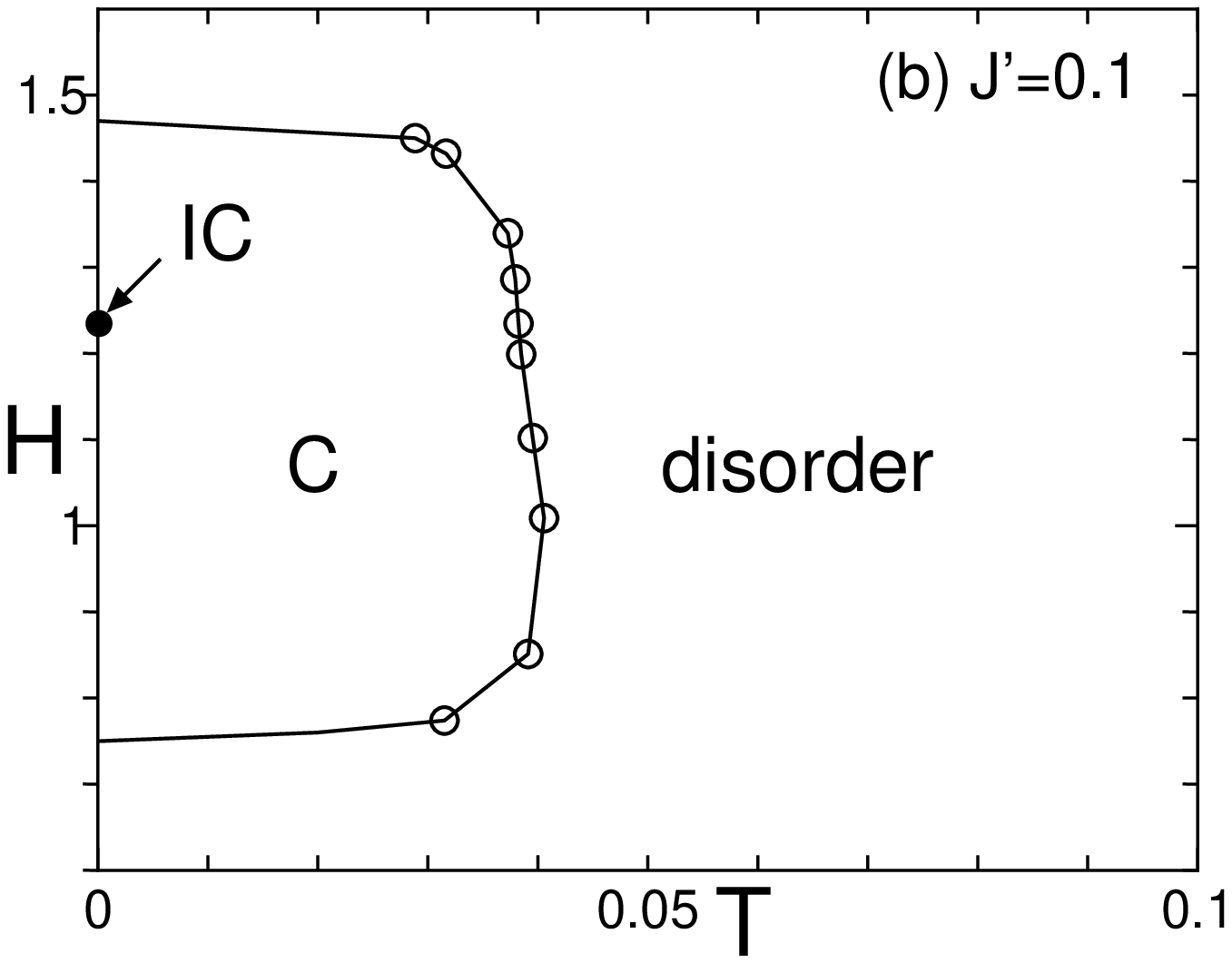}
\includegraphics[width=0.48\linewidth,angle=0]{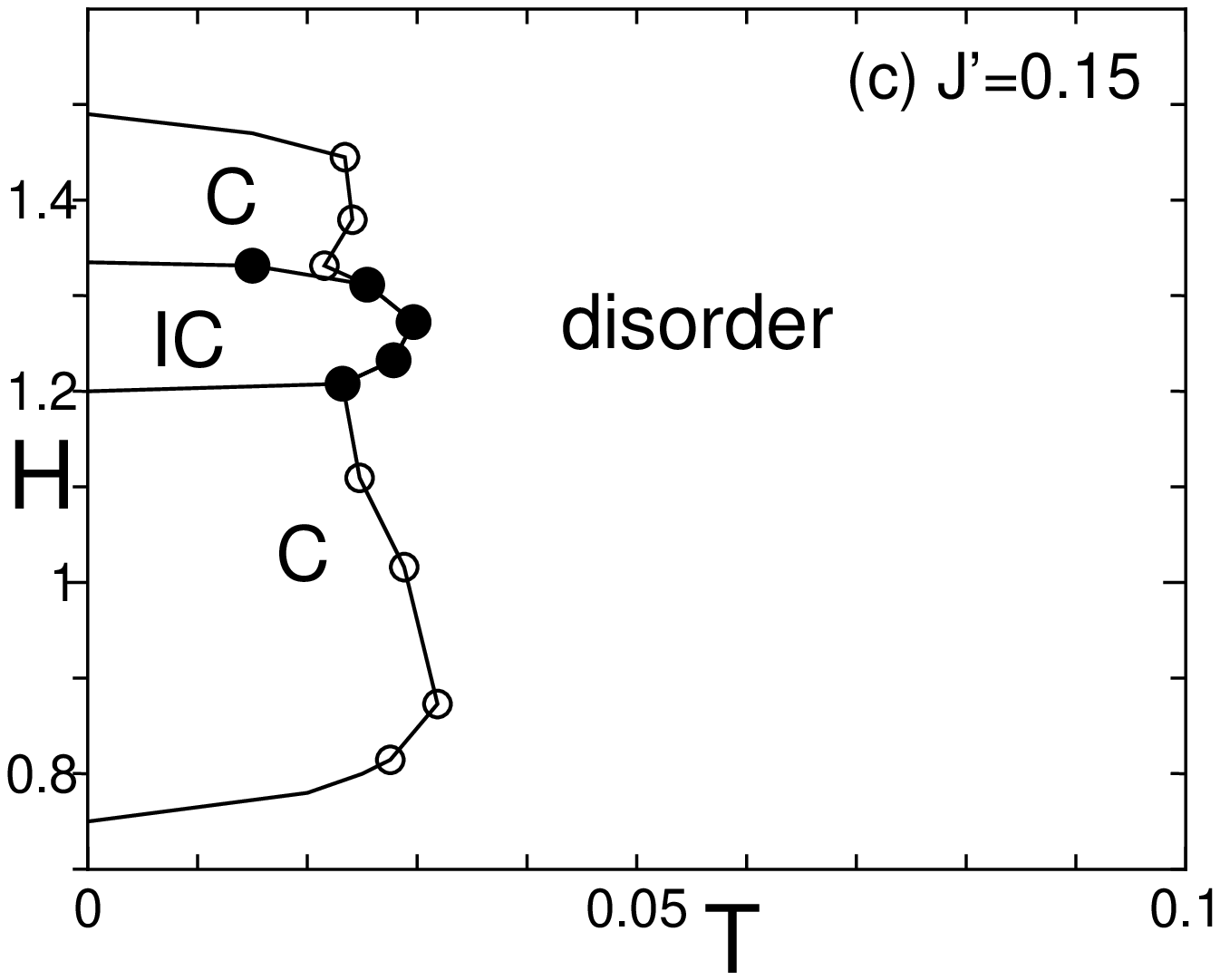}
\caption{
$H-T$ phase diagram for various $J'$  ($J'=0.05$, 0.1 and 0.15).
C and IC denote the commensurate antiferromagnetic and incommensurate order phases, respectively.
Filled circles denote the transition temperature between the IC phase and other phases, and open circles
indicate that between the C phase and the disordered phase.
The interchain interaction is fixed to $zJ_{\rm int}=1/12$.
}\label{fig1}\end{center}\end{figure}

\section{Discussion}

\begin{figure}[btp]
\begin{center}\leavevmode
\includegraphics[width=0.48\linewidth,angle=0]{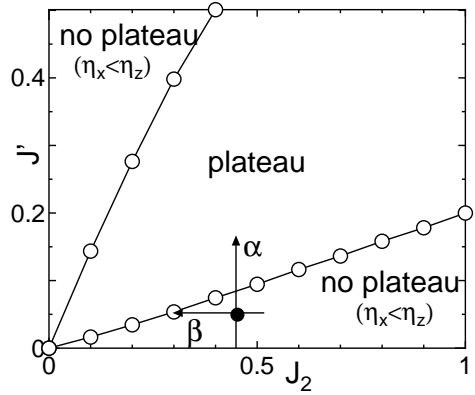}
\caption{
$J_2$-$J'$ phase diagram of the dominant correlations at $m=m_s/2$
 at zero temperature, which is equivalent to the 1/2 plateau phase diagram in reference~\cite{tonegawa}.
The filled circle corresponds to the parameters of F$_5$PNN.
}\label{fig:IC}\end{center}\end{figure}

In the previous work~\cite{maeshima1}, the authors estimated $J'$ of F$_5$PNN as $\sim 0.05$(= 0.3K), based on a qualitative analysis of the magnetization curve.  
This value is slightly smaller than the critical value ($\sim 0.08$) for the $\eta$-inversion or the emergence of the incommensurate order (see figure~\ref{fig:IC}).
We however think that high-pressure experiments can realize the incommensurate phase within realistic experimental situations.  For example, a hydrostatic pressure experiment reveals a drastic change of the bond-alternation ratio in powder samples of F$_5$PNN~\cite{mito}.  In addition, a recent experiment clarifies that specific heat of the power samples has slightly different temperature dependence from that of a single crystal of F$_5$PNN, which can be attributed to an effective pressure caused by an experimentally technical reason~\cite{yoshida3}.
These experimental results strongly indicate that the properties of F$_5$PNN are quite sensitive to the pressure.
Thus it is worthwhile to discuss precisely how the incommensurate order phase develops in the $H$-$T$ phase diagram near the phase boundary of the $\eta$-inversion.

As in figure~\ref{fig:IC},  we can readily consider possible two paths such that the couplings of F$_5$PNN exceeds the critical value for the incommensurate order:
increasing the next-nearest neighbour interaction $J'$ (arrow $\alpha$) and/or decreasing the alternating coupling $J_2$ (arrow $\beta$).
Here, we chiefly discuss the $J'$-dependence of the $H$-$T$ phase diagrams along the vertical arrow direction in figure~\ref{fig:IC}, assuming that  $J_{\rm int}$ is independent of the pressure.  
We can then expect that the features of $H$-$T$ phase diagrams should change as figure~\ref{fig1} (a) $\to$  figure~\ref{fig1} (b) $\to$  figure~\ref{fig1} (c).  
If $J'$ is sufficiently enhanced by the pressure, the direct phase transition from the disordered phase to the IC phase would be observed in a middle range of the magnetic field as in  the case (c). 
As mentioned in the previous section,  however, a notable point is that, even in the case (b), we may find the deformation of the phase boundary between the disordered phase and the C phase.
From the experimental view, such a deformation of the phase boundary can be an important signal of a development of the IC order induced by the frustration, though it may be difficult to capture the IC phase at a very low temperature.

\begin{figure}[btp]
\begin{center}\leavevmode
\includegraphics[width=0.48\linewidth,angle=0]{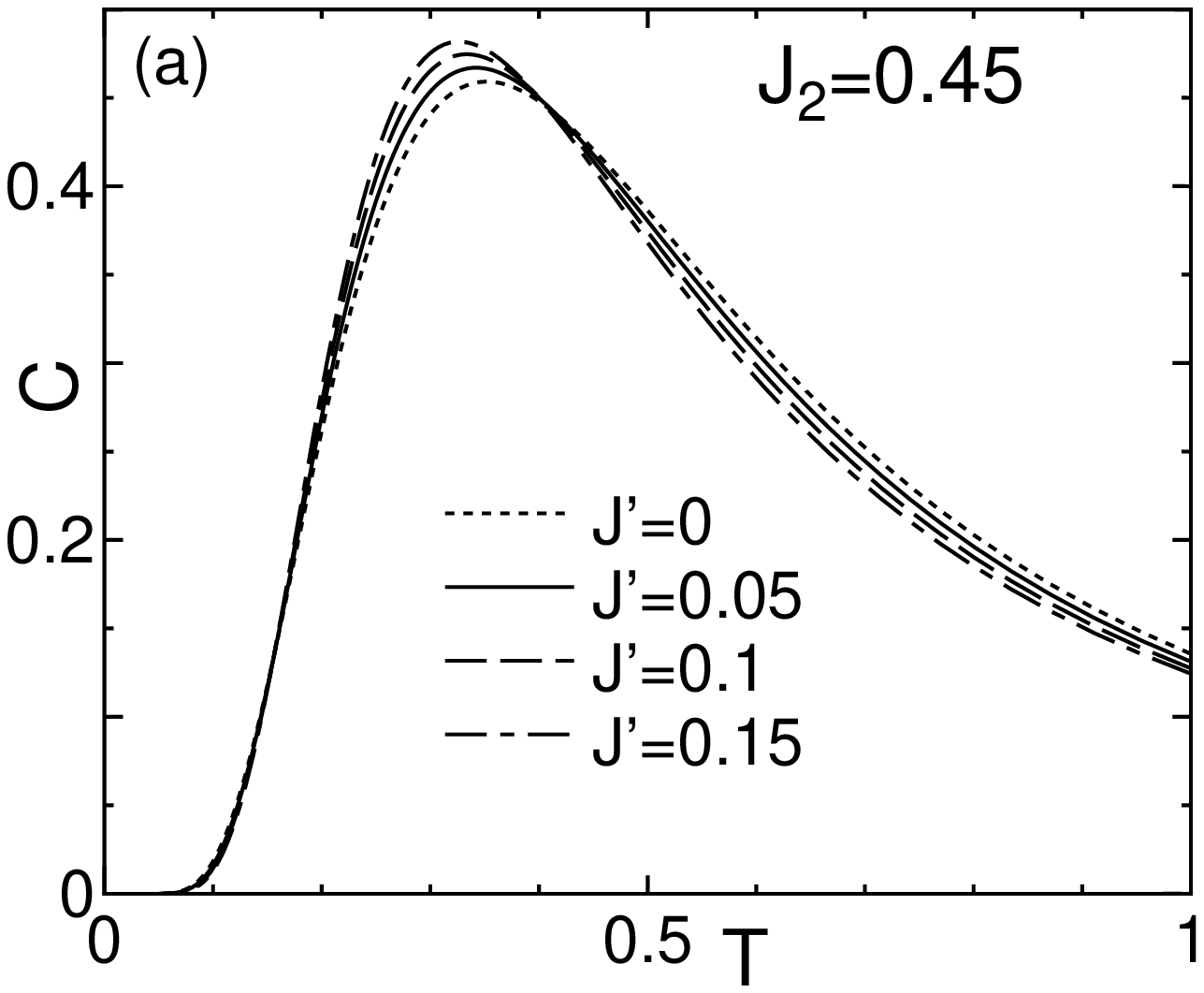}
\includegraphics[width=0.48\linewidth,angle=0]{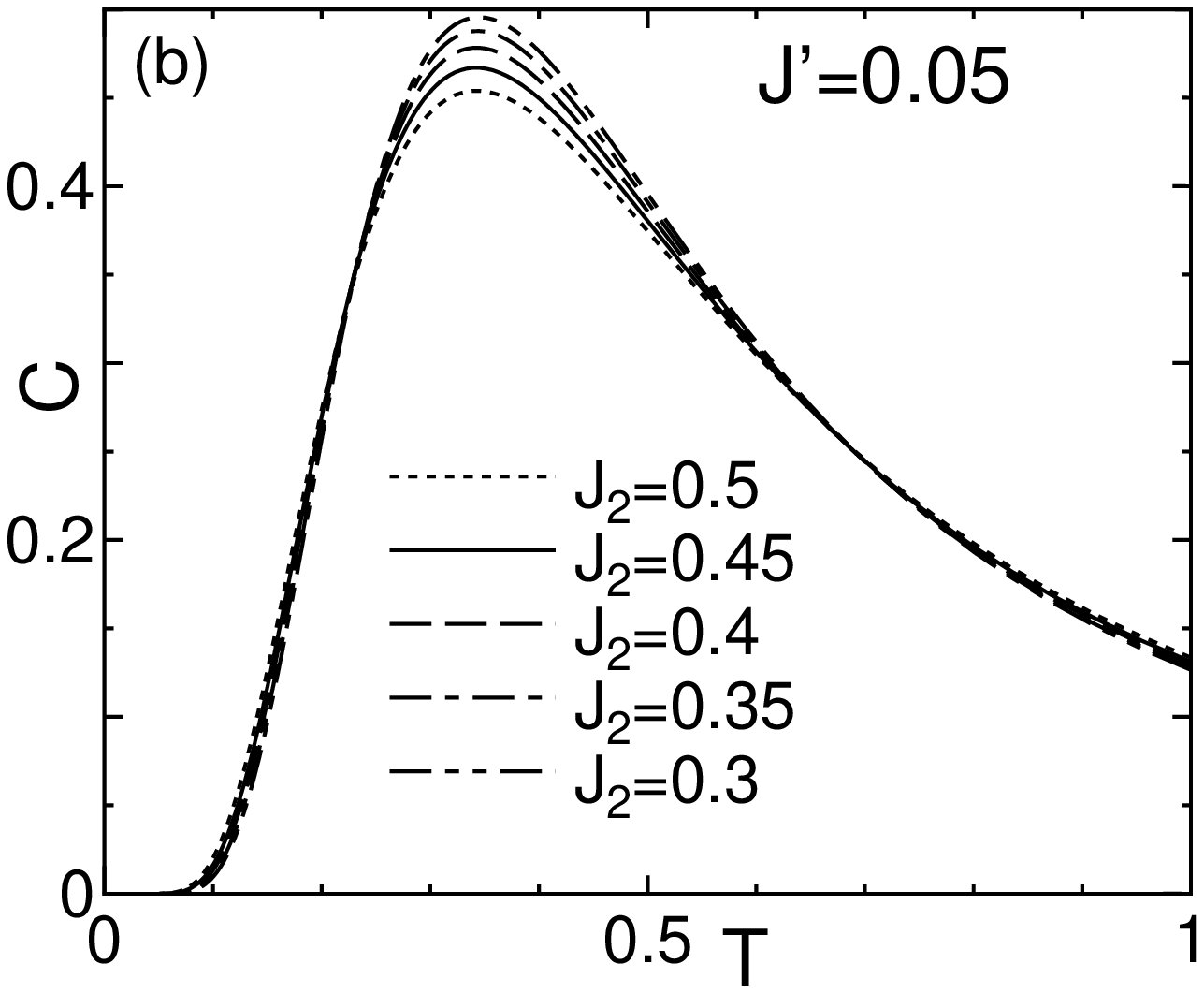}
\caption{
Temperature dependence of the specific heat for various values of exchange parameters.  Each figure corresponds to the case where the exchange interactions are varied along the arrows $\alpha$ and $\beta$ in figure~\ref{fig:IC}.
}\label{fig:capa}\end{center}\end{figure}

Next, we discuss how the changes of the exchange interactions can reflect on a measurement of the specific heat at {\it the zero magnetic field}, which is quite helpful in analysis of the experimental results, because it is not an easy work to obtain a complete $H$-$T$ phase diagram in actual experiments~\cite{yoshida1,yoshida2}.
 Figure~\ref{fig:capa} shows temperature dependence of the specific heat for various values of $J'$ and $J_2$ along the arrows in figure~\ref{fig:IC}.
When the frustrating interaction $J'$ increases along the $\alpha$ direction, it is found that the pronounced peak around $T=0.3$ sharpens.
We note that this can be attributed to the increase of the density of state caused by the narrowing of the $S=1$ magnon band width~\cite{totsuka}.
On the other hand, such a band narrowing also occurs, as  $J_2$ is reduced along the $\beta$ direction. We can then see that the peak sharpens as well. 
We therefore  think that a careful analysis of the shape of the specific heat peak at $H=0$  provides an important information for the enhancement of the  IC correlation in the intermediate magnetic field region, although the pressure effect subtly influences the both of $J'$ and $J_2$ in the actual situation. 

\begin{figure}[btp]
\begin{center}\leavevmode
\includegraphics[width=0.48\linewidth,angle=0]{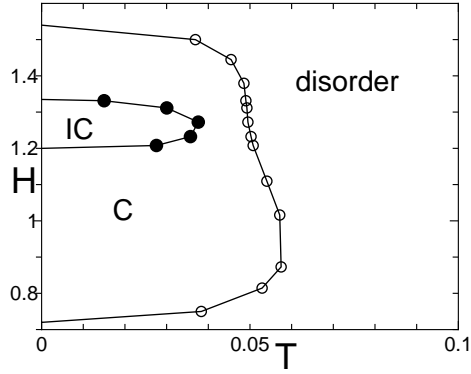}
\caption{
$H$-$T$ phase diagram in case of large $J'$ and large $J_{\rm int}$ ($J'=0.15$ and $zJ_{\rm int}=1/6$). 
}\label{fig2}\end{center}\end{figure}

Here, we should make a comment on the pressure effect for the {\it interchain} interaction.  If $J_{\rm int}$ is also enhanced by the pressure, the direct phase transition from the disordered phase to the IC phase may be suppressed within the interchain mean-field analysis as shown in figure~\ref{fig2}, where the interchain interaction is assumed to be unfrustrating. 
 Nevertheless,  the precise analysis of the crystal structure~\cite{hosokoshi} of F$_5$PNN has indicated that  the interchain couplings are also frustrating, which would effectively decrease value of $J_{\rm int}$ in the mean-field approximation.
In order to reveal the role of the frustrating interchain interaction, of course, a precise analysis beyond the mean-field level is clearly required.

\begin{figure}[btp]
\begin{center}\leavevmode
\includegraphics[width=0.52\linewidth,angle=0]{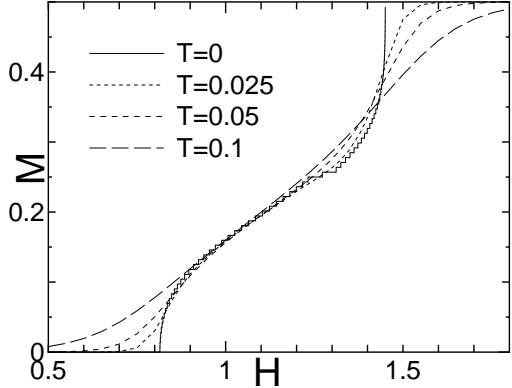}
\caption{
Magnetization curves for the realistic parameters 
$J_2=0.45$ and $J'=0.15$ at several temperatures. 
}\label{fig4}\end{center}\end{figure}

Finally, we want to discuss a magnetization plateau at the half of the full-moment.
As in figure~\ref{fig:IC},  the 1/2 plateau is closely related to the mechanism of the $\eta$-inversion; the criterion for the $\eta$-inversion is equivalent to the one for the plateau formation~\cite{maeshima1}.
The level spectroscopy analysis~\cite{tonegawa} on the model (\ref{ham}) has indicated the 1/2 plateau should appear at half the saturated magnetization for $J_2=0.45$ and $J'=0.15$. 
However,  we think that it may be difficult to observed the clear evidence of the 1/2 plateau of F$_5$PNN even with a high pressure experiment. 
This is because the width of the plateau is still very narrow for these couplings, and thus the thermal fluctuation  smears out the plateau even at temperatures where we can observe the magnetization.
In figure~\ref{fig4}, we show the magnetization curves for  $J_2=0.45$ and $J'=0.15$ obtained by the finite temperature DMRG,  illustrating that the plateau is easily smeared out by the thermal effect. 
We think that, as for F$_5$PNN, it is rather hopeful to verify the presence of the IC order  than to make a direct observation of the  1/2 plateau.

To summarize, we have discussed field-induced IC order for  the bond-alternating spin chain with the next-nearest interaction, which is relevant to the high pressure experiments of F$_5$PNN.
In particular, we have analyzed how the various physical quantities behaves near the phase boundary, and then concluded that the precise analysis for the phase boundary and the zero-field specific heat can detect a signal of the field-induced IC order. Further experimental studies on F$_5$PNN are quite interesting.
In addition, it is also a challenging issue to search another spin chains exhibiting the field-induced incommensurate order.
A recently observed new phase of NDMAP in the magnetic field might be a possible candidate~\cite{takano}.

The field-induced commensurate order in a quantum spin system is often described  as the Bose Einstein condensation (BEC) of a magnon gas.
The present IC order may be viewed as a result of a ``charge order'' of the magnon.
As mentioned in reference~\cite{maeshima2} the transition between the field-induced incommensurate order and  the usual commensurate antiferromagnetic one is of the first order.
Then we can expect the coexistence of the two orders at the phase boundary.
Recently, such a coexistence of the two different orders is actually observed in solid $^4$He, which is called ``super-solid'' state~\cite{supersolid}.
A connection to such an exotic state is also an interesting problem  both from experimental and theoretical viewpoints.


\ack
We wish to thank Profs. I. Affleck, N. Prokof'ev, S. Suga, T. Goto, 
Drs. K. Izumi, T. Suzuki and Y. Yoshida for fruitful 
discussions. 
This work is partly supported by Grants-in-Aid for Scientific Research (B) (No.17340100, No.14340099), (C) (No.14540329, No.16540332, No.17540317), and  Grant-in-Aid for Scientific Research on priority Areas ``High Field Spin Science in 100T''(No.451) from the Ministry of Education, Culture, Sports, Science and Technology of Japan.



\section*{References}

\begin{thebibliography}{99}

\bibitem{sakai1} Sakai T and Takahashi M 1991 {\it Phys. Rev.} B {\bf 43} 13383

\bibitem{sakai11} Sakai T and Takahashi M 1991 {\it J. Phys. Soc. Japan} {\bf 60} 3615

\bibitem{sakai2} Sakai T 2000 {\it Phys. Rev.} B {\bf 62} R9240

\bibitem{honda}
Honda Z, Asakawa H and Katsumata K 1998 {\it Phys. Rev. Lett.} {\bf 81} 2566

\bibitem{tanaka}
Oosawa A, Ishii M and Tanaka H 1999 {\it J. Phys.: Condens. Matter} {\bf 11} 265

\bibitem{nikuni}
Nikuni T, Oshikawa M, Oosawa A and Tanaka H 2000 {\it Phys. Rev. Lett.} {\bf 84} 5868

\bibitem{suga1}
Usami M and Suga S 1998 {\it Phys. Rev.} B {\bf 58} 14401

\bibitem{suga2}
Haga N and Suga S 2000 {\it J. Phys. Soc. Japan} {\bf 69} 2431

\bibitem{suga3}
Suzuki T and Suga S 2004 {\it Phys. Rev.} B {\bf 70} 054419

\bibitem{chitra}
Chitra R and Giamarchi T 1997 {\it Phys. Rev.} B {\bf 55} 5816


\bibitem{mila}
Mila F 1998 {\it Eur. Phys. J.} B {\bf 6} 201

\bibitem{izumi}
Izumi K, Goto T, Hosokoshi Y and Boucher J -P 2003 {\it Physica} B {\bf 329-333} 1191

\bibitem{goto}
Goto T 2004 Private communications

\bibitem{maeshima1}
Maeshima N, Okunishi K, Okamoto K and Sakai T 2004 {\it Phys. Rev. Lett.} {\bf 93} 127203

\bibitem{yoshida1}
Yoshida Y, Yurue K, Mitoh M, Kawae T, Hosokoshi Y, Inoue K, Kinoshita M and Takeda K 2003 {\it Physica} B {\bf 329-333} 979

\bibitem{yoshida2}
Yoshida Y, Tateiwa N, Mito M, Kawae T, Takeda K, Hosokoshi Y and Inoue K 2005 {\it Phys. Rev. Lett.} {\bf 94} 037203

\bibitem{hosokoshi}
Takahashi M, Hosokoshi Y, Nakano H, Goto T, Takahashi M and Kinoshita M 1997 {\it Mol. Cryst. Liq. Cryst.} {\bf 306} 111

\bibitem{scalapino} Scalapino D J, Imry Y, and Pincus P 1975 {\it Phys. Rev.} B {\bf 11} 2042

\bibitem{comment} For $\eta_x<\eta_z$, $C\perp>C_\parallel$ is usually satisfied, where the frustration effect is not significant.

\bibitem{maeshima2} Maeshima N, Okunishi K, Okamoto K, Sakai T and Yonemitsu K 2005 {\it J. Phys. Soc. Japan} {\bf 74} Suppl.  63

\bibitem{wang} Wang X and Xiang T 1997 {\it Phys. Rev.} B {\bf 56} 5061

\bibitem{shibata} Shibata N 1997 {\it J. Phys. Soc. Japan} {\bf 66} 2221


\bibitem{mito}
Mito M, Kawae T, Hosokoshi Y, Inoue K, Kinoshita M and Takeda K 1997 {\it \SSC} {\bf 111} 607

\bibitem{yoshida3}
Yoshida Y, Kawae T, Takeda K, Hosokoshi Y and Inoue K 2005 {\it The Physical Society of Japan 2005 Autumn Meeting } (The Physical Society of Japan). In an actual experiment, a powder sample was mixed with grease.
This grease often causes an effective pressure to the sample even at the ambient pressure.



\bibitem{totsuka} Totsuka K 1998 {\it Phys. Rev.} B {\bf 57} 3454

\bibitem{tonegawa} Tonegawa T, Hikihara T, Okamoto K and Kaburagi M 2001 {\it Physica} B {\bf 294-295} 39

\bibitem{takano} Tsujii H, Honda Z, Andraka B, Katsumata K and Takano Y 2005 {\it Phys. Rev.} B {\bf 71} 014426

\bibitem{supersolid} Kim E and Chan M W H 2004 {\it Nature} {\bf 427} 225



\end{thebibliography}
\end{document}